\documentclass{article}
\usepackage{spconf,amsmath,graphicx}
\usepackage{amssymb}
\usepackage{comment}
\usepackage{pdfpages}
\usepackage{caption}

\DeclareMathOperator*{\argmin}{arg\,min}

\title{ONE SHOT LEARNING FOR SPEECH SEPARATION}
\name{Yuan-Kuei Wu$^1$$^\star$ \qquad Kuan-Po Huang$^2$$^\star$ \qquad Yu Tsao$^3$ \qquad Hung-yi Lee$^4$
\thanks{$^\star$The two first authors made equal contributions.}}
\address{
$^{14}$Graduate Institute of Communication Engineering, National Taiwan University\\
$^{2}$Graduate Institute of Computer Science and Information Engineering, National Taiwan University\\
$^3$Research Center for Information Technology Innovation, Academia Sinica\\
$^{124}$\{f07942100, r09922005, hungyilee\}@ntu.edu.tw, $^3$yu.tsao@citi.sinica.edu.tw}
\begin{document}
\ninept
\maketitle
\begin{abstract}
Despite the recent success of speech separation models, 
they fail to separate sources properly while facing different sets of people or noisy environments. To tackle this problem, we proposed to apply meta-learning to the speech separation task. We aimed to find a meta-initialization model, which can quickly adapt to new speakers by seeing only one mixture generated by those people. In this paper, we use model-agnostic meta-learning(MAML) algorithm and almost no inner loop(ANIL) algorithm in Conv-TasNet to achieve this goal. The experiment results show that our model can adapt not only to a new set of speakers but also noisy environments. Furthermore, we found out that the encoder and decoder serve as the feature-reuse layers, while the separator is the task-specific module.
\end{abstract}
\section{Introduction}
\label{sec:intro}
Speech separation (SS) is an important research topic and has been widely studied. Due to its outstanding nonlinear modeling capability, deep-learning algorithms have been used as core models in the state-of-the-art SS systems \cite{convtasnet, dprnn, wavesplit, voice_unknown_speakernum}. Although these deep-learning-based SS approaches have been shown to obtain revolutionary improvements over traditional SS methods, there are still several challenges to further address. 
A notable one is regarding the generalizability. When the speakers for the mixture signals are different, or the acoustic environments are different in training and testing phases, the SS performance may degrade. Since SS is often used as pre-processing in speech-related applications, an imperfect SS performance may result in unsatisfactory speech recognition, speaker identification, and audio classification tasks.

\begin{figure}[h]
\centering
\includegraphics[width=.45\textwidth]{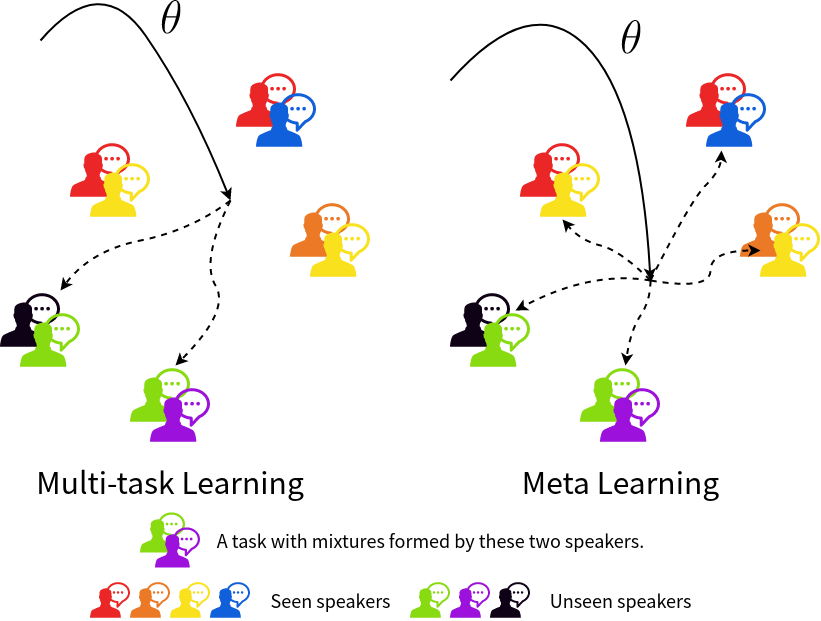}
\caption{Illustration of multi-task learning and meta-learning on speech separation. The solid line represents the learning process of pretraining. The dashed line represents a fast adaptation to unseen speakers.}
\end{figure}

A feasible approach to improve the generalizability of the SS system is to establish a gigantic model, which is trained from mixtures covering different speakers and acoustic environments. Nevertheless, it is challenging to build such a model in two aspects. First, there are always new mixtures of speakers and recording environments, and it is difficult to list all of the combinations exhaustively. Second, collecting labeled data (mixtures and the corresponding single speakers and acoustic environments) is expensive. Therefore, preparing an SS model with a good ability to learn and adapt to new mixtures quickly based on a few examples is a more suitable alternative.

Meta-learning, or the learning-to-learn algorithm, gains great attention in the machine learning field recently. It is different from multi-task learning, which aims to find a model that performs well in all training tasks. The goal of meta-learning is to learn an initialized model that is capable of adapting to or generalizing well on new tasks that are not involved in the training phase. Meta-learning has been applied in computer vision research and achieved great success in image classification\cite{sun2019meta, rusu2018meta, snell2017prototypical, vinyals2016matching}. Meanwhile, several natural language and speech processing tasks also adopt meta-learning and attain promising results, such as neural machine translation\cite{gu2018meta}, dialogue generation\cite{mi2019meta}, text classification\cite{xu2018lifelong}, word sense disambiguation\cite{holla2020learning}, and so on, speaker adaptive training\cite{klejch2018learning},  speech-to-intent classification\cite{s2i_reptile}, code-switched speech recognition\cite{winata2020meta}, and speech recognition\cite{hsu2020meta}.

In this work, we used the MAML algorithm \cite{finn2017model} and the ANIL algorithm \cite{raghu2019rapid} to build an SS system. Among the various meta-learning criteria \cite{snell2017prototypical, finn2017model, raghu2019rapid, sung2018learning}, we believe that these two methods are more suitable for the SS task due to the following properties. First, the model-agnostic property allows them to apply on any models trained with any specific gradient descent methods and loss functions. Second, additional modules are not required, which will not increase the overall system complexity. In summary, the major contribution of the present work is twofold:
\begin{enumerate}
    \item To our best knowledge, this is the first method that applies meta-learning to the SS task. 
    \item Experimental results confirm the effectiveness of the proposed SS system to quickly adapt to different datasets, confirming the system's generalizability has been notably improved when encountering new testing scenarios.
\end{enumerate}

\section{Proposed Methods}
\label{sec:methods}
\subsection{Speech separation model}
\label{ssec:convtasnet}
For the single channel SS task, we want to estimate $C$ sources $\mathbf{s}_1(t), \mathbf{s}_2(t), \dots ,\mathbf{s}_C(t) \in \mathbb{R}^{T}$ from a mixture $\mathbf{x}(t) \in \mathbb{R}^{T}$.
In this study, we used Conv-TasNet\cite{convtasnet} as the SS model. Conv-Tasnet is a fully-convolutional time-domain audio separation network, which consists of an encoder, a separator, and a decoder. First, the encoder maps the mixture $\mathbf{x}(t)$ to a high-dimensional representation $\mathbf{h} \in \mathbb{R}^{D \times T^{'}}$ through applying a 1-D convolutional transformation.
\begin{equation}
    \mathbf{h} = \text{encoder}(\mathbf{x}(t))
\end{equation}
Then, the separator estimates $C$ masks $\mathbf{m}_i  \in \mathbb{R}^{D \times T^{'}}, i = 1, \dots, C$ and generates separated features $\mathbf{d}_i  \in \mathbb{R}^{D \times T^{'}}$.
\begin{equation}
    \mathbf{m}_i = \text{separator}(\mathbf{h})
\end{equation}
\begin{equation}
    \mathbf{d}_i = \mathbf{h} \odot \mathbf{m}_i 
\end{equation}
Finally, the estimated source signals  $\mathbf{\hat{s}}_i \in \mathbb{R}^{1 \times T}$ are obtained by applying transposed 1-D convolutional transformation on the separated features in the decoder.
\begin{equation}
    \mathbf{\hat{s}}_i = \text{decoder}(\mathbf{d}_i)
\end{equation}
The objective function used to train the Conv-TasNet is the scale-invariant source-to-noise ratio (SI-SNR), as shown below:
\begin{equation}
    \label{eqn:sisnr}
    \begin{cases}
        \mathbf{s}_{\text{target}} = \frac{\mathbf{\hat{s}}\cdot \mathbf{s}}{\|\mathbf{s}\|^2} \mathbf{s}\\
        \mbox{SI-SNR} = 10\log_{10}\frac{\|\mathbf{s}_{\text{target}}\|^2}{\|\mathbf{\hat{s}} - \mathbf{s}_{\text{target}}\|^2}
    \end{cases}
\end{equation}
where $\mathbf{s}$ is the ground truth source and $\mathbf{\hat{s}}$ is the estimated signal, and $\|\mathbf{s}\|^2$ is the signal power. $\mathbf{s}_{\text{target}}$ is the projection from $\mathbf{\hat{s}}$ to $\mathbf{s}$. An utterance-level permutation invariant training(uPIT) \cite{uPIT} was used during the loss calculation process to overcome the source permutation problem.
\subsection{Meta-learning tasks for speech separation}
\label{ssec:metass}
In real-world scenarios, the mixtures may be from a different set of speakers. Assume that a scenario includes $n$ speakers $\mathcal{P} = \{p_1, p_2, \dots, p_n\}$, and $\mathcal{A}^i = \{\mathbf{s}^i_j\}$ denotes speech signals from the speaker $p_i$. We first select a subset of speakers $\mathcal{Q} \subseteq \mathcal{P}$. By mixing the speech signals between different speakers in $\mathcal{Q}$ with various SNR (signal-to-noise ratio) levels, we can prepare a source task $T^s_i$. This task is considered as seen when preparing the SS models. The rest speakers and the corresponding utterances are then used to form the target task $T^t_j$, which is assumed unavailable when training the SS models. It is clear that the two tasks contain different speakers, and the SS model trained by $T^s_i$ may not perform well when testing on $T^t_i$.

\subsection{MAML}
The main concept of meta-learning is to prepare a model trained on the source tasks, $\mathcal{T}_{\text{source}} = \{T^s_1, T^s_2, \dots, T^s_n\}$, and the model has good capability to  quickly adapt to the target tasks $\mathcal{T}_\text{target} = \{T^t_1, T^t_2, \dots, T^t_m\}$. The overall process can be divided into two phases, the meta-learning phase, and the fine-tuning phase.
\subsubsection{Meta-learning phase}
 The meta-learning phase tries to find a set of well-initialized parameters for the model, which is easy to be adapted to the target domain. To this end, the algorithm can be divided into two optimization loops, namely inner loop and outer loop. We first draw a batch $\{ \tau_1, \dots , \tau_B\}$ with a batch size $B$ from the distribution of source tasks $P_{\mathcal{T}_{\text{source}}}$. For each task $\tau_b$, we split the data into two sets, support set $\tau^{\text{sup}}_b$ and query set $\tau^{\text{que}}_b$. For the one-shot learning setting, only one mixture is used in the support set.
 
 In the inner loop, we first conduct the task-specific learning process and perform the inner loop separately for each task. For each task $\tau_b$, the learning process can be formulated as finding the parameters that minimize the SI-SNR loss with uPIT over the support set $\tau^{\text{sup}}_b$,
\begin{equation}
    \theta_b = \arg\min_{\theta} \mathcal{L}_{\tau^{\text{sup}}_b}(\theta)
\end{equation}
We calculate $\theta_b$ by using one or multiple gradient descent steps. For simplicity of notation, we formulated it in one gradient descent step with meta-learning rate $\alpha$ as below.
\begin{equation}
\label{eq:adaptation}
    \theta_b \leftarrow \theta - \alpha \nabla_{\theta}\mathcal{L}_{\tau^{\text{sup}}_b}(\theta)
\end{equation}

In the outer loop, we utilize the task-specific models, which are computed by the inner loop, to update the initialized parameters for the generalized model. We define the meta loss function over the query set $\tau^{\text{que}}_b$ as,
\begin{equation}
    \mathcal{L}_{\text{meta}}(\theta) = \sum_{b=1}^{B} \mathcal{L}_{\tau^{\text{que}}_b}(\theta_b)
\end{equation}
The ultimate goal is to find the generalized model, which has the ability to quickly adapt to each task. This can be formulated using the meta loss function. 
\begin{equation}
    \theta^\star = \argmin_{\theta} \mathcal{L}_{\text{meta}}(\theta)
\end{equation}
Finally, we perform gradient descent to update the model parameters.
\begin{equation}
    \theta \leftarrow \theta - \beta \nabla_{\theta}\sum_{b=1}^{B} \mathcal{L}_{\tau^{\text{que}}_b}(\theta_b)
\end{equation}
\subsubsection{Fine-tuning phase}
During the fine-tuning phase, we intend to adapt the model to the target task. In MAML, the adaptation procedure in the fine-tuning phase is the same as that in the meta-learning stage. In the one-shot learning setup, only one data sample $d_j$ in the target task $T^t_j$ is available for adaptation. We obtain the target task-specific model $\theta_j$ by performing gradient descent on the well-initialized model $\theta$ (obtained from the meta-learning phase). The overall procedure is:
\begin{equation}
    \label{eq:finetune}
    \theta_j \leftarrow \theta - \alpha \nabla_{\theta}\mathcal{L}(d_j, \theta)
\end{equation}

\subsection{ANIL}
ANIL, as a simplified version of MAML, provides similar performance on computational efficiency. MAML is featured as \emph{rapid learning}, while ANIL is featured as \emph{feature reuse}. In rapid learning, the meta-initialization in the outer loop learns parameters that are favorable for fast adaptation. Thus, a huge number of parameters and representational changes take place in the inner loop. In feature reuse, the meta-initialization already contains useful information that can be reused. Therefore, only little changes are required in the inner loop. Assume the model $\theta$ can be split into two parts, generalized feature extraction layers, and task-specific layers. Let $\omega$ be the meta initialization parameters of task-specific layers. Since the generalized feature extraction layers learned the shared representations for all tasks, they will not be updated during adaptation. Compared to MAML, ANIL only updates the task-specific layers in the inner loop. Thus the adaptation procedure, which is different from Eq. \ref{eq:adaptation}, can be formulated as,
\begin{equation}
\label{eq:anil_adaptation}
    \omega_b \leftarrow \omega - \alpha \nabla_{\omega}\mathcal{L}_{\tau^{\text{sup}}_b}(\omega)
\end{equation}
For the same reason, we only update the task-specific layers in the fine-tuning phase. Therefore, we modified Eq. \ref{eq:finetune} to
\begin{equation}
    \label{eq:afinetune}
    \omega_j \leftarrow \omega - \alpha \nabla_{\omega}\mathcal{L}(d_j, \omega).
\end{equation}
Finally, with fixed feature extraction layers and updated  task-specific layers, we can then obtain a SS model for the target task. 

\section{Experiment Setups}
In the experiment, we used Conv-TasNet with the configuration that yields the best performance in \cite{convtasnet} as the core SS model. We focused on one-shot learning in the two-speaker SS task in the experiment. Also, we prepared the datasets for meta-learning SS (metaSS), which are called WSJ0-2mix-meta, Libri-2mix-meta, and VCTK-2mix-meta.
\label{sec:experiments}
\subsection{Dataset}
\label{ssec:dataset}
\begin{figure}[t]
  \centering
  \includegraphics[width=.4\textwidth]{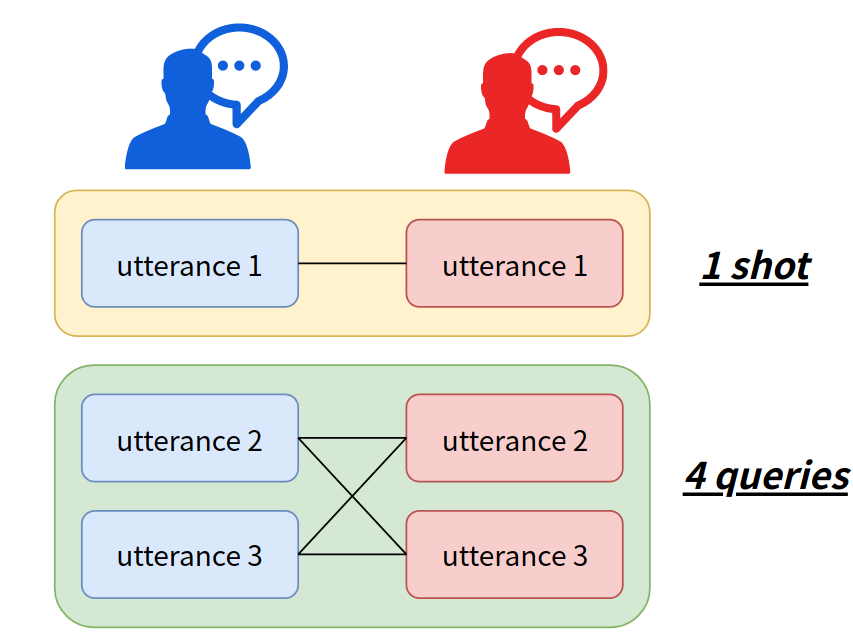}
  \caption{We prepared nine mixtures in a task to form the support set and the query set. In the one-shot scenario, there is only one mixture in the support set. Considering that a source speech cannot appear in both support and query sets, there are only four mixtures in the query set.}
  \label{fig:data_diagram}
\end{figure}

We used WSJ0-2mix-meta as our training set, Libri-2mix-meta , and VCTK-2mix-meta dataset as our testing sets. The WSJ0-2mix-meta, Libri-2mix-meta and VCTK-2mix-meta datasets are generated from Wall Street Journal dataset (WSJ0)\cite{wsj0}, LibriSpeech\cite{librispeech} and VCTK corpus\cite{vctk}, respectively. All mixtures in these meta SS datasets are generated by the following process. First, we selected three utterances of each speaker from the source dataset.
Then, we selected two speakers to form a task. In each task, we mixed the utterances from these two speakers at a random SNR level between 0 dB and 5 dB and resampled the mixture to 8 kHz. Thus, there will be $3 \times 3 = 9$ mixtures in one task. The example of how to form the support and query set is illustrated in Fig. \ref{fig:data_diagram}. The WSJ0 corpus has 101 speakers in si\_tr\_s. All of them are included to form WSJ0-2mix-meta. Hence, a total of ${101 \choose 2} = 5050$ tasks are generated. These meta tasks are then split into 4050 tasks (54.7 hours) and 1000 tasks (13.5 hours) for the training and develop sets, respectively. 

For the testing sets, the Libri-2mix-meta dataset is constructed from the LibriSpeech test-clean set. There are 40 speakers with clean utterances in the LibriSpeech test-clean set. We sampled 14 speakers to construct 91 tasks (1 hour) for testing. The VCTK-2mix-meta dataset was created from the VCTK dataset. There are 109 English speakers with different accents in the VCTK dataset. We also randomly chose 14 speakers to form 91 tasks (0.7 hours) in VCTK-2mix-meta.

We used the noise profiles in the MUSAN corpus \cite{musan_noise} as the noise sources. The dataset included various technical and non-technical noise profiles, such as fax machine noises, DTMF tones, and dial tones. Furthermore, ambient sounds such as rain, paper rustling, animal noises, are also included. The noise sources were mixed with the SS data in Libri-2mix-meta and VCTK-2mix-meta between SNR 10 dB and 15 dB SNR levels for testing.

 

\subsection{Multi-task \& transfer learning}
\label{ssec:baseline}
To deal with the low resource problem in the target domain, there are several works \cite{winata2019code, tong2017investigation, zoph2016transfer} using transfer learning algorithms. The main idea is to pretrain and leverage the knowledge learned from a high resource dataset, and then perform transfer learning to adapt the pretrained model to match the low resource data. We build our baseline model in this scenario. In the pretraining phase, we trained the SS model on the WSJ0-2mix-meta dataset. The data of all tasks are used to train the SS model.

In the transfer learning phase, the model is updated over an example in the support set and evaluated on four examples in the query set for each task. We implemented and compared the performance using three kinds of adaptation methods. First, we adopt the same process as Eq. \ref{eq:finetune} and denoted it as "m"(stands for MAML) in the experiments. Second, the task-specific layers are updated, which is the same as Eq. \ref{eq:afinetune}. We tested the performance of using either the separator (denoted as "a\_s") or the autoencoder (encoder \& decoder, denoted as "a\_c") as the task-specific layer.


\subsection{Meta-learning}
\label{ssec:meta-learninig}
In the experiments, we analyzed two meta-learning methods, MAML and ANIL. Before meta-learning training, we used the pretrained model mentioned in section \ref{ssec:baseline} as our initial model. We then trained the model with the meta-learning method on the WSJ0-2mix-meta training set for 25 epochs. Then, we picked the best model by evaluating on the WSJ0-2mix-meta development set. For ANIL, we considered two cases for the task-specific module, which we updated in the inner loop and during the fine-tuning phase. For ANIL\_s, we viewed the separator as the task-specific module; for ANIL\_c, we viewed the autoencoder as the task-specific module. The fine-tuning and evaluation process is the same as section \ref{subsec:multi}. The performance is measured in 
SI-SNRi \cite{vincent2006performance}.


\section{Results}
\label{sec:results}

\subsection{Multi-task \& meta-learning}
\label{subsec:multi}
\begin{table}[ht]
\centering
\begin{tabular}{cccccccc}
\toprule
   &method                       & p.t.                   & f.t.         & libri        & vctk         & libri\_n     & vctk\_n      \\ \midrule
(1)&                             & best                   & -            & 8.97         & 5.32         & 7.00         & 4.35         \\  
(2)&                             & best                   & m            & 8.80         & 4.90          & 6.88         & 3.98         \\ 
(3)&                             & best                   & a\_s         & 9.06         & 5.51         & 7.54         & 4.57         \\  
(4)&                             & best                   & a\_c         & 8.99         & 5.12         & 7.47         & 4.56         \\ 
(5)&                             & half                   & -            & 8.35         & 5.08         & 6.37         & 4.28         \\ 
(6)&\multirow{-6}{*}{Multitask}  & half                   & m            & 8.53         & 5.16         & 6.56         & 4.39         \\ \midrule
(7)&                             & best                   & m            &\textbf{9.84} & 7.76         & 7.56         & 5.99         \\ 
(8)&                             & half                   & m            & 9.55         & 7.94         &\textbf{7.59} & 6.38         \\  
(9)&\multirow{-3}{*}{MAML}       & -                      & m            & 9.38         &\textbf{8.62} & 7.54         &\textbf{7.18} \\ \bottomrule
\end{tabular}
\caption{Evaluation results of multi-task learning and MAML with different pretrained epochs and fine-tuning methods on Libri-2mix-meta and VCTK-2mix-meta with and without noise involvement.}
\label{table:results}
\end{table}

\begin{figure}[t]
    \centering
    \subfloat{
        \centering
        \begin{tikzpicture}
        \begin{axis}[
            width=.45\textwidth,
            height=5.9cm,
            xlabel={$\alpha$},
            ylabel={Si-SNRi},
            xmin=0.5, xmax=10.5,
            ymin=-3, ymax=11,
            xtick={1,...,10},
            xticklabels={1e-6, 5e-6, 1e-5, 5e-5, 1e-4, 5e-4, 1e-3, 5e-3, 1e-2, 5e-2},
            x tick label style={rotate=40,anchor=east},
            ytick={-3,...,10},
            yticklabels={, -2, , 0, , 2, , 4, , 6, , 8, , 10},
            legend style={nodes={scale=0.8, transform shape}}, 
            legend pos=north east,
            ymajorgrids=true,
            grid style=dashed,
            label style={font=\small},
            tick label style={font=\small} 
        ]
        
        \addplot[
            color=blue,
            mark=square,
            mark options={scale=0.5}
            ]
            coordinates {
                (10,-0.84)
                (9,3.14)
                (8,4.385)
                (7,6.39)
                (6,6.85)
                (5,6.51)
                (4,6.78)
                (3,6.64)
                (2,6.735)
                (1,6.61)
                };
        \addplot[
            color=black,
            mark=square,
            mark options={scale=0.5}
            ]
            coordinates {
                (10,-0.36)
                (9,3.37)
                (8,4.815)
                (7,6.645)
                (6,6.92)
                (5,6.815)
                (4,6.74)
                (3,6.62)
                (2,6.77)
                (1,6.585)
                };
        \addplot[
            color=red,
            mark=square,
            mark options={scale=0.5}
            ]
            coordinates {
                (10,2.645)
                (9,5.42)
                (8,5.98)
                (7,6.685)
                (6,6.765)
                (5,6.73)
                (4,6.65)
                (3,6.735)
                (2,6.65)
                (1,6.68)
                };        
            \legend{$m$, $a\_s$, $a\_c$}
            
        \end{axis}
        \end{tikzpicture}
    }
    \caption{For each fine-tuning method, we evaluate the performance by adjusting the learning rate $\alpha$ in the range of $10^{-6}$ to 0.05. We took the average of the SI-SNRi scores in all testing datasets.}
    \label{fig:fastlr}
\end{figure}
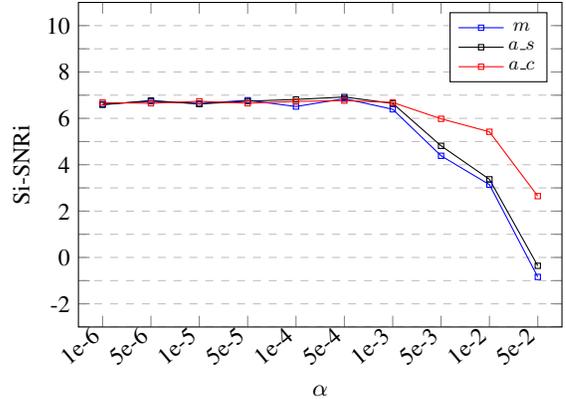

In this experiment, we analyzed the meta-learning method (the MAML method was reported as a representative here), where the multi-task learning is also reported for comparison. In multi-task learning, we first pretrained our model on the WSJ0-2mix-meta training set for 100 epochs. During training, we identified the best model by evaluating the models on the WSJ0-2mix-meta development set. The best model was trained for 84 epochs, and this model will be used in the following experiments. Apart from the best model, we also chose the model trained for 50 epochs for the experiment, which we refer to as "half" in Table \ref{table:results}. In the table, the pretrained models by multi-task training were used for models (7) and (8). We also tested the fine-tune learning rate(referred as $\alpha$ in Eq. \ref{eq:finetune} and \ref{eq:afinetune}) of multi-task learning models in the range between $10^{-6}$ and $0.05$. The results are shown in Fig. \ref{fig:fastlr}. To obtain optimal baseline models, a learning rate that achieved the best performance was used. We also set the learning rate to 0.01 for the meta-learning models, which is consistent to the training stage.

For MAML, model (7), which is based on the the best pretrained model, obtained the best performance on Libri-2mix-meta; model (8), which is based on the half pretrained model, achieved the highest $\text{SI-SNRi}$ score on Libri-2mix-meta with noise; moreover, model (9) yields the best score on the VCTK-2mix-meta both with and without noise. It is clear that all the best results are from the models trained based on the MAML method. Thus, we can conclude that meta-learning models outperform multi-task learning not only on the test sets of unseen speakers but also SS with noise involved. 

By further comparing the results of models {(1), (5)} and {(2), (6)}, the performance drops in most cases while using the half pretrained model. However, we can see a different result from MAML by comparing between models (7) and (8). Though we used the half pretrained model rather than the best pretrained model, the performances are still better in most of the cases. This suggests that we do not need to pretrain our models using all training epochs in MAML in most of the cases.

From results in models {(2), (3), (4)}, we can see the effect of different fine-tuning methods for multi-task learning. Model (3) performs better than models (2) and (4), indicating that fine-tuning on the separator module outperforms fine-tuning on the whole model and the autoencoder. This suggests that the layers in the separator play a more important role in the task-specific layers in the SS model. Comparing these methods to model (1), which is not fine-tuned after multi-task learning, models (3) and (4) perform better while model (2) suffered from performance degradation. The results show that fine-tuning on task-specific layers is more effective than multi-task learning without fine-tuning.

\subsection{Feature reuse \& rapid learning}
\begin{table}[h]
\centering
\begin{tabular}{cccccccc}
\toprule
   &method                       & p.t.                   & f.t.         & libri        & vctk         & libri\_n     & vctk\_n      \\ \midrule
(1)&                             & best                   & m            &\textbf{9.84} & 7.76         & 7.56         & 5.99         \\ 
(2)&\multirow{-2}{*}{MAML}       & -                      & m            & 9.38         &\textbf{8.62} & 7.54         &\textbf{7.18} \\ \midrule
(3)&                             & best                   & a\_s         & 9.67         & 7.92         &\textbf{7.64} & 6.17         \\  
(4)&\multirow{-2}{*}{ANIL\_s}    & -                      & a\_s         & 9.48         & 7.57         & 7.53         & 6.16         \\ \midrule
(5)&ANIL\_c                      & best                   & a\_c         & 8.89	        & 6.52	       & 7.03	      & 5.33         \\ \bottomrule
\end{tabular}
			
\caption{The evaluation results for different meta-learning methods with different pretraining and fine-tuning methods on two testing sets with and without noise.}
\label{table:meta_results}
\end{table}
Table \ref{table:meta_results} shows the result of the complete meta-learning results by both MAML and ANIL. By comparing models (3) and (5), ANIL\_s performs better in all the datasets, showing that the separator is the task-specific module, while the encoder and the decoder are the modules that generalize reused features. However, we can see that updating and fine-tune on the whole model is the best way in the meta-learning methods by comparing models (1), (2), (3), and (4). Thus, rapid learning is the best way for the SS model to adapt to new speakers and noisy environments.



\section{Conclusion}
\label{sec:conclusion}
In this paper, we proposed meta-learning approaches on SS with the aim to quickly adapt to new situations. Experimental results confirmed that the proposed method outperforms the multi-task learning method on new speakers and noisy conditions. Also, we analyzed different methods and found out that the separator plays a more important role for the task-specific module in SS. Because the model-agnostic method is used in this study, we adopted Conv-TasNet as the SS model throughout our experiments. In future work, we will test the meta-learning methods on other SS models.
\section{Acknowledgement}We thank to National Center for High-performance Computing (NCHC) for providing computational and storage resources.






\bibliographystyle{IEEEbib}
\bibliography{strings,refs}

\begin{thebibliography}{10}

\bibitem{convtasnet}
Yi~Luo and Nima Mesgarani,
\newblock ``Conv-tasnet: Surpassing ideal time--frequency magnitude masking for
  speech separation,''
\newblock {\em IEEE/ACM transactions on audio, speech, and language
  processing}, vol. 27, no. 8, pp. 1256--1266, 2019.

\bibitem{dprnn}
Yi~Luo, Zhuo Chen, and Takuya Yoshioka,
\newblock ``Dual-path rnn: efficient long sequence modeling for time-domain
  single-channel speech separation,''
\newblock in {\em ICASSP 2020-2020 IEEE International Conference on Acoustics,
  Speech and Signal Processing (ICASSP)}. IEEE, 2020, pp. 46--50.

\bibitem{wavesplit}
Neil Zeghidour and David Grangier,
\newblock ``Wavesplit: End-to-end speech separation by speaker clustering,''
\newblock {\em arXiv preprint arXiv:2002.08933}, 2020.

\bibitem{voice_unknown_speakernum}
Eliya Nachmani, Yossi Adi, and Lior Wolf,
\newblock ``Voice separation with an unknown number of multiple speakers,''
\newblock {\em arXiv preprint arXiv:2003.01531}, 2020.

\bibitem{sun2019meta}
Qianru Sun, Yaoyao Liu, Tat-Seng Chua, and Bernt Schiele,
\newblock ``Meta-transfer learning for few-shot learning,''
\newblock in {\em Proceedings of the IEEE conference on computer vision and
  pattern recognition}, 2019, pp. 403--412.

\bibitem{rusu2018meta}
Andrei~A Rusu, Dushyant Rao, Jakub Sygnowski, Oriol Vinyals, Razvan Pascanu,
  Simon Osindero, and Raia Hadsell,
\newblock ``Meta-learning with latent embedding optimization,''
\newblock {\em arXiv preprint arXiv:1807.05960}, 2018.

\bibitem{snell2017prototypical}
Jake Snell, Kevin Swersky, and Richard Zemel,
\newblock ``Prototypical networks for few-shot learning,''
\newblock in {\em Advances in neural information processing systems}, 2017, pp.
  4077--4087.

\bibitem{vinyals2016matching}
Oriol Vinyals, Charles Blundell, Timothy Lillicrap, Daan Wierstra, et~al.,
\newblock ``Matching networks for one shot learning,''
\newblock in {\em Advances in neural information processing systems}, 2016, pp.
  3630--3638.

\bibitem{gu2018meta}
Jiatao Gu, Yong Wang, Yun Chen, Kyunghyun Cho, and Victor~OK Li,
\newblock ``Meta-learning for low-resource neural machine translation,''
\newblock {\em arXiv preprint arXiv:1808.08437}, 2018.

\bibitem{mi2019meta}
Fei Mi, Minlie Huang, Jiyong Zhang, and Boi Faltings,
\newblock ``Meta-learning for low-resource natural language generation in
  task-oriented dialogue systems,''
\newblock {\em arXiv preprint arXiv:1905.05644}, 2019.

\bibitem{xu2018lifelong}
Hu~Xu, Bing Liu, Lei Shu, and Philip~S Yu,
\newblock ``Lifelong domain word embedding via meta-learning,''
\newblock {\em arXiv preprint arXiv:1805.09991}, 2018.

\bibitem{holla2020learning}
Nithin Holla, Pushkar Mishra, Helen Yannakoudakis, and Ekaterina Shutova,
\newblock ``Learning to learn to disambiguate: Meta-learning for few-shot word
  sense disambiguation,''
\newblock {\em arXiv preprint arXiv:2004.14355}, 2020.

\bibitem{klejch2018learning}
Ond{\v{r}}ej Klejch, Joachim Fainberg, and Peter Bell,
\newblock ``Learning to adapt: a meta-learning approach for speaker
  adaptation,''
\newblock {\em arXiv preprint arXiv:1808.10239}, 2018.

\bibitem{s2i_reptile}
Yusheng Tian and Philip~John Gorinski,
\newblock ``Improving end-to-end speech-to-intent classification with
  reptile,''
\newblock {\em arXiv preprint arXiv:2008.01994}, 2020.

\bibitem{winata2020meta}
Genta~Indra Winata, Samuel Cahyawijaya, Zhaojiang Lin, Zihan Liu, Peng Xu, and
  Pascale Fung,
\newblock ``Meta-transfer learning for code-switched speech recognition,''
\newblock {\em arXiv preprint arXiv:2004.14228}, 2020.

\bibitem{hsu2020meta}
Jui-Yang Hsu, Yuan-Jui Chen, and Hung-yi Lee,
\newblock ``Meta learning for end-to-end low-resource speech recognition,''
\newblock in {\em ICASSP 2020-2020 IEEE International Conference on Acoustics,
  Speech and Signal Processing (ICASSP)}. IEEE, 2020, pp. 7844--7848.

\bibitem{finn2017model}
Chelsea Finn, Pieter Abbeel, and Sergey Levine,
\newblock ``Model-agnostic meta-learning for fast adaptation of deep
  networks,''
\newblock {\em arXiv preprint arXiv:1703.03400}, 2017.

\bibitem{raghu2019rapid}
Aniruddh Raghu, Maithra Raghu, Samy Bengio, and Oriol Vinyals,
\newblock ``Rapid learning or feature reuse? towards understanding the
  effectiveness of maml,''
\newblock {\em arXiv preprint arXiv:1909.09157}, 2019.

\bibitem{sung2018learning}
Flood Sung, Yongxin Yang, Li~Zhang, Tao Xiang, Philip~HS Torr, and Timothy~M
  Hospedales,
\newblock ``Learning to compare: Relation network for few-shot learning,''
\newblock in {\em Proceedings of the IEEE Conference on Computer Vision and
  Pattern Recognition}, 2018, pp. 1199--1208.

\bibitem{uPIT}
Morten Kolb{\ae}k, Dong Yu, Zheng-Hua Tan, and Jesper Jensen,
\newblock ``Multitalker speech separation with utterance-level permutation
  invariant training of deep recurrent neural networks,''
\newblock {\em IEEE/ACM Transactions on Audio, Speech, and Language
  Processing}, vol. 25, no. 10, pp. 1901--1913, 2017.

\bibitem{wsj0}
Douglas~B Paul and Janet Baker,
\newblock ``The design for the wall street journal-based csr corpus,''
\newblock in {\em Speech and Natural Language: Proceedings of a Workshop Held
  at Harriman, New York, February 23-26, 1992}, 1992.

\bibitem{librispeech}
Vassil Panayotov, Guoguo Chen, Daniel Povey, and Sanjeev Khudanpur,
\newblock ``Librispeech: an asr corpus based on public domain audio books,''
\newblock in {\em 2015 IEEE International Conference on Acoustics, Speech and
  Signal Processing (ICASSP)}. IEEE, 2015, pp. 5206--5210.

\bibitem{vctk}
Christophe Veaux, Junichi Yamagishi, Kirsten MacDonald, et~al.,
\newblock ``Superseded-cstr vctk corpus: English multi-speaker corpus for cstr
  voice cloning toolkit,''
\newblock 2016.

\bibitem{musan_noise}
David Snyder, Guoguo Chen, and Daniel Povey,
\newblock ``Musan: A music, speech, and noise corpus,''
\newblock {\em arXiv preprint arXiv:1510.08484}, 2015.

\bibitem{winata2019code}
Genta~Indra Winata, Andrea Madotto, Chien-Sheng Wu, and Pascale Fung,
\newblock ``Code-switched language models using neural based synthetic data
  from parallel sentences,''
\newblock {\em arXiv preprint arXiv:1909.08582}, 2019.

\bibitem{tong2017investigation}
Sibo Tong, Philip~N Garner, and Herv{\'e} Bourlard,
\newblock ``An investigation of deep neural networks for multilingual speech
  recognition training and adaptation,''
\newblock in {\em Proc. of INTERSPEECH}, 2017, number CONF.

\bibitem{zoph2016transfer}
Barret Zoph, Deniz Yuret, Jonathan May, and Kevin Knight,
\newblock ``Transfer learning for low-resource neural machine translation,''
\newblock {\em arXiv preprint arXiv:1604.02201}, 2016.

\bibitem{vincent2006performance}
Emmanuel Vincent, R{\'e}mi Gribonval, and C{\'e}dric F{\'e}votte,
\newblock ``Performance measurement in blind audio source separation,''
\newblock {\em IEEE transactions on audio, speech, and language processing},
  vol. 14, no. 4, pp. 1462--1469, 2006.

\end{thebibliography}

\end{document}